\documentclass[preprint,12pt]{elsarticle}

\usepackage{lineno,hyperref}
\modulolinenumbers[5]

\usepackage{amssymb}
\usepackage{amssymb}
\usepackage{graphicx}
\usepackage{amsmath,amsfonts,amssymb}
\usepackage{epsfig,makecell,float}
\usepackage{setspace,mathrsfs}
\usepackage{tocloft}
\usepackage{textcomp}
\usepackage{multirow,indentfirst,times,color}
\usepackage[caption=false,font=footnotesize]{subfig}
\usepackage{booktabs}
\usepackage{threeparttable}
\usepackage{amsthm}
\usepackage{extarrows}

\usepackage[titletoc]{appendix}
\usepackage{epstopdf}
\usepackage{natbib}
\biboptions{numbers,sort&compress}

\newtheorem{defn}{Definition}
\newtheorem{remark}{Remark}

\begin{document}

\begin{frontmatter}

\title{Discrete Linear Canonical Transform on Graphs
}

\author{Yu Zhang$^{a,b}$}
\author{Bing-Zhao Li$^{a,b}$\corref{mycorrespondingauthor}}
\cortext[mycorrespondingauthor]{Corresponding author}\ead{li\_bingzhao@bit.edu.cn}

\address{$^{a}$School of Mathematics and Statistics, Beijing Institute of Technology, Beijing 100081, China}
\address{$^{b}$Beijing Key Laboratory on MCAACI, Beijing Institute of Technology, Beijing 100081, China}


\begin{abstract}
	With the wide application of spectral and algebraic theory in discrete signal processing techniques in the field of graph signal processing, an increasing number of signal processing methods have been proposed, such as the graph Fourier transform, graph wavelet transform and windowed graph Fourier transform. In this paper, we propose and design the definition of the discrete linear canonical transform on graphs (GLCT), which is an extension of the discrete linear canonical transform (DLCT), just as the graph Fourier transform (GFT) is an extension of the discrete Fourier transform (DFT). First, based on the centrality and scalability of the DLCT eigendecomposition approach, the definition of GLCT is proposed by combining graph chirp-Fourier transform, graph scale transform and graph fractional Fourier transform. Second, we derive and discuss the properties and special cases of GLCT. Finally, some GLCT examples of the graph signals are given to illustrate the improvement of the transformation.
	
\end{abstract}


\end{frontmatter}


\section{Introduction}
\label{Intro}
In conventional signal processing, the signal domain is determined by a set of spatial measurement points at equal time intervals or on a uniform grid. However, with the advancement of information and communication, actual signals may not be related to the temporal or spatial dimension, and they exhibit various forms of irregularities, including social irregularities, economical irregularities, irregularities in transportation, irregularities in energy, irregularities in sensors, and irregularities in biological networks. These high-dimensional and massive signals require new processing techniques, which eventually developed the field of graph signal processing \cite{1}, \cite{2}. Signal processing on graphs extends traditional discrete signal processing (DSP) to signals with underlying complex and irregular structures and achieves the transition from classical discrete signal processing to graph signal processing by graphically modeling and vertex indexing the underlying structure of the signals. Related works include graph-based transformation \cite{3,4,5,6,7}, filtering \cite{3}, \cite{8,9}, sampling and interpolation \cite{10,11,12}, reconstruction and restoration \cite{13,14,15}, denoising \cite{12}, \cite{16}, uncertainty principle on graphs \cite{17}, etc.

There are two basic signal processing frameworks on the graph based on spectral and algebraic methods. The first method is derived from spectrogram theory \cite{1}. The graph signal is expanded into the eigenfunction of the Laplacian operator to define the graph Fourier transform, and the corresponding spectrum is represented by the eigenvalues. Since standard graph Laplacian matrices must be symmetric and positive semidefinite, this method is limited to undirected graphs. The second is based on discrete signal processing on graphs (DSPG) \cite{2,3}, which is derived from algebraic signal processing (ASP). The graph Fourier transform expands the signal into the eigenvectors of the adjacency matrix and defines the spectrum by the corresponding eigenvalues. Since the adjacency matrix can be asymmetric, the second method can be applied to arbitrary graphs. However, this transformation requires the information of the entire graph signal, and the conversion process from the vertex domain to the graph frequency domain cannot be obtained. To solve problems such as nonstationary graph signals, the windowed graph Fourier transform \cite{6}, the fractional Fourier transform on the graph \cite{18} and the windowed fractional Fourier transform on the graph \cite{19} have been proposed, which further expands the methods and application prospects of graphs. However, the above approaches still have the problems of insufficient degrees of freedom, inadequate flexibility, and underutilized parameters. Therefore, finding new graph signal processing methods is currently a popular topic.

To address the above problems, we apply the linear canonical transform (LCT) to graph signal processing. The LCT was proposed in the 1970s  \cite{20} and later introduced into the field of signal processing in 2001 \cite{21}. The linear canonical transforms (LCTs) generalize Fourier transforms, fractional Fourier transforms, Fresnel transforms, and scaling operations and have three free parameters. Before the implementation of the DLCT into the field of graph signals, the central discrete dilated Hermite function (CDDHF) \cite{22, 23} was used as the basis for the eigenfunctions of the fractional Fourier transform (FrFT) and scaling operations to implement the DLCT. The introduction of LCT into the field of graph signal processing can effectively solve the lack of the degrees of freedom and flexibility, expand the method of graph signal processing, and unify the graph Fourier transform, graph fractional Fourier transform, and graph scale transform. In this paper, a definition of the discrete linear canonical transform on graphs (GLCT) is proposed and consolidated by the following steps.

Summary of the paper: In Section II, the theories of graph signal and discrete linear canonical transformation on which this study is based are briefly reviewed. In Section III, the definition of GLCT and several properties are presented. In Section IV, examples of the graph signals under the GLCT framework are given to illustrate the improvements of the transform. Finally, in Section V, the paper is concluded.

\section{Preliminaries}
\label{Preli}
\subsection{Graph Fourier transform}
The graph Fourier transform is defined using algebraic methods, and some of the most basic concepts and definitions are reviewed below. A detailed introduction to the theory can be found in \cite{2, 3}.

We consider a dataset with $N$ elements, where the associated information of some of the data elements is known through dependency. This relationship is represented by the graph $\mathcal{G}=(\mathcal{V},\mathbf{A})$, where $\mathcal{V}={v_0,\hdots,v_{N-1}} $ is the set of vertices, and $\mathbf{A}$ is the weighted adjacency matrix of the graph. Each data element is indexed by a node $v_n$, and each weighted edge $\mathbf{A}_{n,m}\in \mathbb{C}$ from $v_m$ to $v_n$ represents the relation of the $m$th data element to the $n$th data element. For undirected graphs, there is $\mathbf{A}_{n,m} = \mathbf{A}_{m,n}$, that is, $\mathbf{A}$ is symmetric.
The dataset is called the graph signal, defined as the following: 
\begin{equation}
\begin{aligned}\mathbf{s} :\mathcal{V} &\rightarrow \mathbb{C} ,\\ v_{n}&\mapsto s_{n},\end{aligned} 
\end{equation}
where $\mathbb{C} $ is the set of complex numbers. It can be written as graph signals as vectors
\[
\mathbf{s} = \begin{bmatrix}
s_0 & s_1& \hdots & s_{N-1}
\end{bmatrix}^T \in \mathbb{C}^N.
\]

For simplicity of the discussion, it is assumed that $\mathbf{A}$ is diagonalizable and its eigendecomposition is
\begin{equation}
\mathbf{A} = \mathbf{V\Lambda V^{-1}} ,
\end{equation}
where the columns $v_n$ of the matrix  $\mathbf{V} = \begin{bmatrix}
v_0 & v_1& \hdots & v_{N-1}\end{bmatrix} \in \mathbb{C}^{N\times N}$ are the eigenvectors of $\mathbf{A}$, and $\mathbf{\Lambda} \in \mathbb{C}^{N\times N}$ is the diagonal matrix of the distinct eigenvalues $\lambda_0,~\lambda_1,\dots,\lambda_{N-1}$ of $\mathbf{A}$ .The eigenvalues are the graph frequencies that form the spectrum of the graph. 
If $\mathbf{A}$ is not diagonalizable, Jordan decomposition\cite{2,3} into generalized eigenvectors is uesd. 

\begin{defn}
The graph Fourier transform of a graph signal can be defined as
\begin{equation}
	\hat{\mathbf{s}}=\mathbf{Fs}:=\mathbf{V^{-1}s},
\end{equation}

where $\mathbf{F} =\mathbf{V}^{-1}$ is the graph Fourier transform matrix, $
\hat{\mathbf{s}} = \begin{bmatrix}	\hat{s}_0 & \hat{s}_1& \hdots & \hat{s}_{N-1}\end{bmatrix}^T$. 

The inverse graph Fourier transform is defined as
\begin{equation}
	\mathbf{s}=\mathbf{F^{-1}\hat{s}}:=\mathbf{V\hat{s}}.
\end{equation}
\end{defn}

\subsection{ Graph  Fractional Fourier transform}
\label{GFrFT}
To further study graph signals and the fractional Fourier transform, the discrete fractional Fourier transform (DFrFT) is extended to the graph fractional Fourier transform (GFrFT) \cite{18}, \cite{24}, similar to the extension of DFT to GFT.
\begin{defn}
The $a$th fractional graph Fourier transform of the graph signal $s \in \mathbb{C}^{N \times N}$ is defined as
\begin{equation}
	\mathbf{\hat{s}}= \mathcal{F}^a \mathbf{s} := \mathbf{F}^{a}s.
\end{equation}
\end{defn}

The signal $\mathbf{s}$ on the graph $\mathcal{G}$ has a diagonalizable adjacency matrix $\mathbf{A=V\Lambda_{(A)} V}^{-1}$. If the GFT matrix $\mathbf{F}$ is ortho-diagonalized, then

\begin{equation}
\mathbf{F} = \mathbf{V}^{-1} = \mathbf{Q \Lambda Q }^{-1},
\end{equation}
where the matrices $\mathbf{Q}$ and $\mathbf{\Lambda}$ are given by the spectral decomposition of the GFT matrix. Similar to the definition of the $a$th DFrFT matrix, the GFrFT matrix can be given by:
\begin{equation}
\mathbf{\hat{s}} =\mathbf{F}^{a}=\mathbf{Q}\mathbf{\Lambda}^{a} \mathbf{Q}^{-1}\mathbf{s},\  \   a\in \left[ 0,1\right] ,
\end{equation}
and its inverse transform (IGFRFT) is
\begin{equation}
\mathbf{s} =\mathbf{F}^{-a}=\mathbf{Q}\mathbf{\Lambda}^{-a}\mathbf{ Q}^{-1}\mathbf{\hat{s}}.
\end{equation}

\subsection{Linear Canonical Transforms}
\subsubsection{Definition}
The linear canonical transform of the signal $x(t)$ with the parameters$(a,b,c,d)$ is defined as
\begin{equation}
\begin{aligned}L^{\left( a,b,c,d\right)  }\left( x\left( t\right)  \right)  =&\sqrt{\frac{1}{ib} } \cdot \int^{\infty }_{-\infty } \exp \left[ i\pi \left( \frac{d}{b} u^{2}-\frac{2}{b} ut+\frac{a}{b} t^{2}\right)  \right]  \\ &\times x\left( t\right)  \mathrm{d} t\  \  \  \  \  \text{when} \  b\neq 0,\\ L^{\left( a,0,c,d\right)  }\left( x\left( t\right)  \right)  =&\sqrt{d} \cdot \exp \left( i\pi dct^{2}\right)  x\left( d\cdot t\right)  \\ &\times \text{when} \  b=0,\end{aligned} 
\end{equation}
where $L^{\left( a,b,c,d\right)  } $ denotes the LCT operator. The matrix $\mathbf{M}$ entries are $(a,b,c,d)$, which is called an LCT parameter matrix, and $ad-bc=1, \ a,b,c,d\in \mathbb{R}$.

The DLCT can also be decomposed into chirp multiplication(CM), scaling and FrFT:
\begin{equation}
\left[ \begin{matrix}a&b\\ c&d\end{matrix} \right]  =\left[ \begin{matrix}1&0\\ \xi &1\end{matrix} \right]  \left[ \begin{matrix}\sigma &0\\ 0&\sigma^{-1} \end{matrix} \right]  \left[ \begin{matrix}\cos \alpha &\sin \alpha \\ -\sin \alpha &\cos \alpha \end{matrix} \right]  ,\label{DLCT}
\end{equation}
where $\xi$ denotes the CM parameter, $\sigma$ denotes the scaling parameter, and $\alpha$ denotes the FrFT parameter and satisfies the determinant constraint $ad-bc=1$. The parameter relations between $(a,b,c,d)$ and $(\xi,\sigma,\alpha)$ are 
\begin{equation}
\begin{gathered}\xi =\frac{ac+bd}{a^{2}+b^{2}} ,\  \  \  \  \  \sigma =\sqrt{a^{2}+b^{2}} ,\\ \alpha =\cos^{-1} \left( \frac{a}{\sigma } \right)  =\sin^{-1} \left( \frac{b}{\sigma } \right)  .\end{gathered} 
\end{equation}

\subsubsection{Discretization Process}
There are three main ways to define and calculate the DLCT: the direct discretization method \cite{25} based on classical discrete-time Fourier transform ideas and methods, the linear canonical transform decomposition method based on the principle of the FFT algorithm \cite{26, 27, 28}, and the parameter matrix decomposition method \cite{29}. Because a discrete Hermite function with centralization and expandability is required when implementing the DLCT, the approach of eigendecomposition of the CDDHFs is adopted in this paper to achieve the DLCT. Since the LCT matrix is not commutative, it is implemented in the order of DFrFT, scaling transform, and CM \cite{24}, \cite{30}. For a discrete signal $x[n]$, we have
\begin{equation}
L^{\left( \cos \alpha ,\sin \alpha ,-\sin \alpha ,\cos \alpha \right)  }\left( x\left[ n\right]  \right)  =ED_{\alpha }E^{T}x\left[ n\right]  , \label{DFrFT}
\end{equation}
where the parameters $(a,b,c,d)= (\cos \alpha, \sin \alpha, -\sin \alpha, \cos \alpha)$ , $E$ is a square matrix of $N\times N$, and $D_{\alpha}$ is a diagonal matrix. Then, the discrete scale transform can be expressed as
\begin{equation}
L^{\left( \sigma ,0,0,\sigma^{-1} \right)  }\left( x\left[ n\right]  \right)  =E_{\sigma }E^{T}x\left[ n\right]  ,
\end{equation}
where the parameters $(a,b,c,d)=(\sigma,0,0,\sigma^{-1})$, $E_{\sigma}$ is an $N \times N$ square matrix, and $\sigma$ is a constant, when $\sigma>1$, the transformed signal is a dilated signal. Finally, the CM is defined as
\begin{equation}
L^{\left( 1,0,\xi ,1\right)  }\left( x\left[ n\right]  \right) =D_{\xi }x\left[ n\right]  , \label{CM}
\end{equation}
where, the parameters $(a,b,c,d)=(1,0,\xi,1)$, and $D_{\xi}$ is the diagonal chirp matrix.

In summary, combining the discrete FrFT, scaling and CM in Eqs\eqref{DFrFT}–\eqref{CM}, DLCT based on Eq\eqref{DLCT} is proposed. Therefore, the cascaded combination of the three operations is:
\begin{equation}
\begin{aligned}L^{\left( a,b,c,d\right)  }\left( x\left[ n\right]  \right)  &=\left\{ \left( D_{\xi }\right)  \left( E_{\sigma }E^{T}\right)  \left( ED_{\alpha }E^{T}\right)  \right\}  x\left[ n\right]  \\ &=\left\{ D_{\xi }E_{\sigma }D_{\alpha }E^{T}\right\}  x\left[ n\right]  .\end{aligned} \label{new DLCT}
\end{equation}

The DLCT Eq\eqref{new DLCT} is a unitary transformation because the DLCT matrix $D_{\xi }E_{\sigma }D_{\alpha }E^{T}$ is unitary, and the matrix $E^{T}E$ is the identity matrix, since $E$ forms an orthogonal Hermite basis set. Figure 1 is a schematic diagram of the eigendecomposition of the DLCT, where $\hat{x} \left[ n\right]  $ is the frequency in the linear canonical domain.

\begin{figure}[h]
\begin{center}
	\includegraphics[scale=0.6]{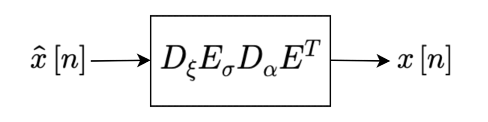}
\end{center}
\vspace*{-16pt}
\caption{Eigendecomposition of the DLCT.}
\vspace*{-3pt}
\end{figure}

\section{ Discrete Linear Canonical Transform on Graphs}
\label{GLCT}
In the field of graph signals, there is little theoretical literature defining chirped signals on graphs, therefore, we propose a definition of the chrip-Fourier transform on graphs \cite{30}. Using the previously proposed definitions of graph scale transforms \cite{31,32} and the fractional Fourier transform on graphs \cite{18}, we finally define a linear canonical transform on graphs.
\subsection{Definition}
This definition generalizes the DLCT in the same way that the GFT generalizes the DFT. Similar to defining the GFrFT, we first assume that the GLCT can be represented in matrix form as
\begin{equation}
\mathbf{\hat{s} } =\mathcal{L}^{(a,b,c,d)} \mathbf{s} :=\mathbf{L}^{(a,b,c,d)} \mathbf{s} .
\end{equation}

Then, we consider the signal $\mathbf{s}$ on the graph $\mathcal{G}$, which has a diagonalizable adjacency matrix $\mathbf{A=V\Lambda V}^{-1}$. It is assumed that the GFT matrix $\mathbf{L}$ can be orthogonally diagonalized.
\begin{equation}
\mathbf{L} = \mathbf{V}^{-1} = \mathbf{Q \Lambda Q^{T} }.
\end{equation}

The GLCT can also be decomposed into three stages of CM, scaling, and FrFT. This method uses the Hermite function (HF) or Hermite-Gaussian function to define the eigenfunctions and eigenvalues. However, the eigenfunctions of the GFrFT and scaling transform derived using discrete HF are not dilated. Therefore, the GLCT is implemented using the CDDHFs as the basis for the eigenfunctions of the FrFT and scaling operations. According to the eigendecomposition of the DLCT in Figure 1, we can deduce the eigendecomposition of the GLCT into the form of Figure 2 by analogy.

\begin{figure}[h]
\begin{center}
	\includegraphics[scale=0.85]{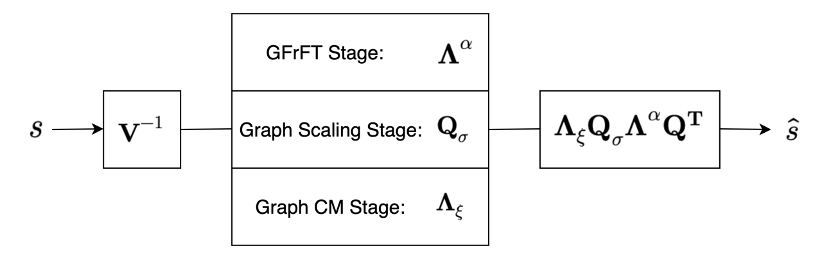}
\end{center}
\vspace*{-15pt}
\caption{Eigendecomposition of the GLCT.}
\vspace*{-3pt}
\end{figure}

To obtain the eigendecomposition matrix shown in Figure 2, we achieve it through the following three steps.
\begin{enumerate}[(1)]
\item GFrFT Stage\\
Since arbitrary LCT matrices are not commutative, the same order as defined for the DFrFT matrices should be implemented first.
\begin{equation}
	\mathbf{L^{\alpha}} = \mathbf{V}^{-1} = \mathbf{Q \Lambda^{\alpha} Q^{T} } \label{eqGFrFT},
\end{equation}
where $\alpha=\frac{\pi }{2} a$ , $a$ is the order of the previous GFrFT in \eqref{GFrFT}.

\item Graph Scaling Stage\\
In classical DSP, using the graph representation of finite, periodic time series in Fig.3 \cite{33}, for which the scaled adjacency matrix is the $N\times N$ circulant matrix $\mathbf{A}=\mathbf{C}$, and weights are
\begin{equation}
	\mathbf{A}_{n,m}=\begin{cases}1,&\text{if} \  n-m=1\  \mathrm{m} \mathrm{o} \mathrm{d} \  N\\ 0,&\text{otherwise} \end{cases} ,
\end{equation}
we can write the scaled adjacency matrix as $\mathbf{A}=\mathbf{C}$, and the shift operation is
\begin{equation}
	\tilde{\mathbf{s} } =\mathbf{C} \mathbf{s} =\mathbf{As}.
\end{equation}

\begin{figure}[h]
	\begin{center}
		\includegraphics{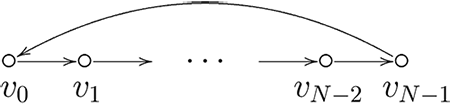}
	\end{center}
	\vspace*{-12pt}
	\caption{Time-series with the length of N represented by a cycle graph.}
	\vspace*{-3pt}
\end{figure}

The process implemented by the linear operator can be expressed as
\begin{equation}
	\left[ x\right]_{n,t}  =p\left[ x\right]_{n,t-1}  +\sum_{m\in E_{n}} q_{n,m}\left[ x\right]_{m,t-1}  ,
\end{equation}
where $q$ is the edge weights of $n$ and $m$, $p$ is an arbitrary weight at $t-1$, and $E_{n}$ is the sum of $\mathcal{G}$ edges. In the case of the scaled adjacency matrix, what each node performs \cite{32} on the signal is

\begin{equation}
	\left[ x\right]_{n,t}  =\frac{1}{\sigma} \sum_{m\in E_{n}} \left[ x\right]_{m,t-1}  .
\end{equation}

In this way,  the scaled adjacency matrix is $\mathbf{A^{\prime }} = \frac{1}{\sigma} \mathbf{A}$ as the graph shift operator. When in an arbitrary graph $\mathcal{G}=(\mathcal{V},\mathbf{A})$, the graph shift operation is:
\begin{equation}
	\tilde{\mathbf{s}}_{n} =\sum^{N-1}_{m=0} \mathbf{A}_{n,m}\mathbf{s}_{m}=\sum_{m\in E_{n}} \mathbf{A}_{n,m}\mathbf{s}_{m}.
\end{equation}
This matrix $\mathbf{A}_{n,m}$ is the scaled adjacency matrix, and $\mathbf{A}_{n,m}=\mathbf{A^{\prime}}$.

We consider that the adjacency matrix $\mathbf{A}$ is first scaled to obtain $\mathbf{A^{\prime }}$, and then the eigenvalues and eigenvectors of $\mathbf{A^{\prime }}$ are obtained. Similar to GFT, the graph spectrum decomposition is
\begin{equation}
	\mathbf{A^{\prime }}=\mathbf{Q_{\sigma }\Lambda^{\prime } Q^{T}_{\sigma }},
\end{equation}
where $\mathbf{Q}_{\sigma }$ and $\mathbf{\Lambda}^{\prime }$ denote the eigenvectors of the matrix 	$\mathbf{A}^{\prime }$ and its corresponding eigenvalues, espectively. $\sigma$ is the scaling parameter.

\item Graph CM Stage\\
Because of its noncommutativity, the CM is performed on the adjacency matrix $\mathbf{A}$ , as 
\begin{equation}
	\begin{aligned}\mathbf{L_{\xi }} =&\mathbf{V}^{-1} =\mathbf{Q\Lambda_{\xi } Q^{T}} \\ =&\mathbf{Q} \begin{bmatrix}\mathrm{\lambda }_{0} &&\\ &\ddots &\\ &&\lambda_{N-1} \end{bmatrix}_{\xi } \mathbf{Q^{T}} =\mathbf{Q} \begin{bmatrix}\mathrm{\lambda }_{\xi_{0} } &&\\ &\ddots &\\ &&\lambda_{\xi_{N-1} } \end{bmatrix} \mathbf{Q^{T}} ,\end{aligned}  \label{GCM}
\end{equation}
where $\xi$ is the CM parameter, and $\mathbf{\Lambda}_{\xi}$ is the eigenvalues of the matrix $\mathbf{A}$ after scaling and CM.

The discrete chirp matrix is defined in terms of the discrete chirp-Fourier transform \cite{30}. The chirp-Fourier transform is closely related to the fractional Fourier transform, where the angle is related to the variable $l$, hence, we define the chirp transform on the graph as:
\begin{equation}
	F\left[ m,n\right]  =\sum^{N-1}_{k=0} u_{k}\left[ m\right]  \exp \left( j\frac{\pi }{2} k\left( \xi\right)  \right)  u_{k}\left[ n\right]  ,
\end{equation}
where $\xi=lk+f,$ $\xi$ is a linear function of $k,$ $l$ and $f$ are constants, and the graph eigenvalues are $\lambda^{\xi }_{k} $.
\end{enumerate}

Based on the above information, we formally provide the definition of the GLCT as follows:
\begin{defn}
We have described and shown the graph FrFT, scaling and CM phases in Eqs\eqref{eqGFrFT}-\eqref{GCM}. The DLCT is the proposed equation-based GLCT. Therefore, it can be expressed as a cascading combination of these three operations:
\begin{equation}
	\mathbf{L}^{\left( a,b,c,d\right)  }=\mathbf{V}^{-1}=\mathbf{\Lambda}_{\xi } \mathbf{Q}_{\sigma }\mathbf{\Lambda}^{\alpha } \mathbf{Q^{T}} ,
\end{equation}
and the definition of the GLCT of the signal $\mathbf{s}$ is
\begin{equation}
	\mathbf{\hat{s}} =\mathcal{L}^{\left( a,b,c,d\right)  }\mathbf{s}:=\mathbf{L}^{\left( a,b,c,d\right)  }\mathbf{s}=\left\{ \mathbf{\Lambda}_{\xi } \mathbf{Q}_{\sigma }\mathbf{\Lambda}^{\alpha } \mathbf{Q^{T}}\right\}  \cdot \mathbf{s} ,
\end{equation}
where the parameter relations between $(a,b,c,d)$ and $(\xi,\sigma,\alpha)$ are 
\begin{equation}
	\begin{gathered}\xi =\frac{ac+bd}{a^{2}+b^{2}} ,\  \  \  \  \  \sigma =\sqrt{a^{2}+b^{2}} ,\\ \alpha =\cos^{-1} \left( \frac{a}{\sigma } \right)  =\sin^{-1} \left( \frac{b}{\sigma } \right)  .\end{gathered} 
\end{equation}
\end{defn}

\begin{defn}
In our GLCT approach, the invertibility can be obtained by additivity as follows
Its inverse transform is
\begin{equation}
	\begin{aligned}\mathbf{L}^{\left( d,-b,-c,a\right)  } \left\{ \mathbf{L}^{\left( a,b,c,d\right)  } \mathbf{s} \right\}  &=\mathbf{s} \\ \  \text{i.e.} ,\  \left[ \begin{matrix}a&b\\ c&d\end{matrix} \right]^{-1}  \left[ \begin{matrix}a&b\\ c&d\end{matrix} \right]  &=\left[ \begin{matrix}1&0\\ 0&1\end{matrix} \right]  ,\end{aligned} 
\end{equation}
where $\mathbf{L}^{\left( d,-b,-c,a\right)  }$ denotes the inverse GLCT (IGLCT) operator of  $\mathbf{L}^{\left( a,b,c,d\right)  }$. Thus, its inverse transform is
\begin{equation}
	\mathbf{s} =\left( \mathcal{L}^{\left( a,b,c,d\right)  } \right)^{-1}  \mathbf{\hat{s} } :=\mathbf{L}^{\left( d,-b,-c,a\right)  } \mathbf{\hat{s} } =\left\{ \mathbf{Q}\mathbf{\Lambda}_{-\alpha } \mathbf{Q^{T}}_{\sigma }\mathbf{\Lambda}_{-\xi } \right\}  \cdot \mathbf{\hat{s} } .
\end{equation}
\end{defn}

\begin{remark}
In particular, when $a=\cos \alpha, b=\sin \alpha, c=-\sin \alpha,$ and $d=\cos \alpha$, the GLCT operator will be the GFRFT operator,
\begin{equation}
	\mathbf{L}^{\left( \cos \alpha ,\sin \alpha ,-\sin \alpha ,\cos \alpha \right)  }=\mathbf{\Lambda}_{0} \mathbf{Q}_{1}\mathbf{\Lambda}_{\alpha } \mathbf{Q^{T}}=\mathbf{IQ}_{1}\mathbf{\Lambda}_{\alpha } \mathbf{Q^{T}}=\mathbf{Q\Lambda}_{\alpha } \mathbf{Q^{T}},
\end{equation}
where $\xi=\frac{-\sin \alpha \cos \alpha +\sin \alpha \cos \alpha }{\cos^{2} \alpha +\sin^{2} \alpha } =0, \sigma =\sqrt{\cos^{2} \alpha +\sin^{2} \alpha } =1.$
\end{remark}
\begin{remark}
When $a=0, b=1, c=-1,$ and $d=0$, the GLCT operator will be the GFT operator,
\begin{equation}
	\mathbf{L}^{\left( 0,1,-1,0\right)  } =\mathbf{\Lambda }_{0} \mathbf{Q}_{1} \mathbf{\Lambda }_{\frac{\pi }{2} } \mathbf{Q^{T}} =\mathbf{IQ} \mathbf{\Lambda }_{\frac{\pi }{2} } \mathbf{Q^{T}} =\mathbf{Q\Lambda }_{\frac{\pi }{2} } \mathbf{Q^{T}} =\mathbf{F} ,
\end{equation}
where $\xi =\frac{-1+1}{1} =0,\sigma =\sqrt{0+1} =1,\alpha =\cos^{-1} \left( 0\right)  =\sin^{-1} \left( 1\right)  =\frac{\pi }{2} $, and $\mathbf{F}$ is the GFT operator.
\end{remark}

\subsection{Properties}
Similar to the GFrFT, we introduce the zero rotation, additivity, and invertibility properties of GLCT, define the convolution operator of GLCT, and derive its translation property.
\begin{enumerate}[(1)]
\item Zero rotation \\
\begin{equation}
	\mathbf{L}^{\left( 1,0,0,1\right)  }=\mathbf{\Lambda}_{0}\mathbf{Q}_{1}\mathbf{\Lambda}_{0} \mathbf{Q^{T}}=\mathbf{IQIQ^{T}}=\mathbf{QQ^{T}}=\mathbf{I},
\end{equation}
where $a=1, b=0, c=0, d=1,$ and calculated $\xi=0, \sigma=1, \alpha=0$.

\item Additivity \\
The additivity of the LCT is expressed as matrix multiplication, and the additivity of the GLCT can be obtained by replacing the matrix as
\begin{equation}
	\mathbf{\Lambda }_{\xi_{3} } \mathbf{Q}_{\sigma_{3} } \mathbf{\Lambda }_{\alpha_{3} } \mathbf{Q^{T}} \left( s\left[ n\right]  \right)  =\mathbf{\Lambda }_{\xi_{2} } \mathbf{Q}_{\sigma_{2} } \mathbf{\Lambda }_{\alpha_{2} } \mathbf{Q^{T}} \left\{ \mathbf{\Lambda }_{\xi_{1} } \mathbf{Q}_{\sigma_{1} } \mathbf{\Lambda }_{\alpha_{1} } \mathbf{Q^{T}} \left( s\left[ n\right]  \right)  \right\}  ,
\end{equation}
where
\begin{equation}
	\left[ \begin{matrix}a_{3}&b_{3}\\ c_{3}&d_{3}\end{matrix} \right]  =\left[ \begin{matrix}a_{2}&b_{2}\\ c_{2}&d_{2}\end{matrix} \right]  \left[ \begin{matrix}a_{1}&b_{1}\\ c_{1}&d_{1}\end{matrix} \right]  .
\end{equation}

\item Invertibility \\
The invertibility of the GLCT means that the IGLCT can be implemented by another GLCT whose parameter matrix equals to the inverse of the forward transform matrix, and the invertibility can be obtained by the above additivity:
\begin{equation}
	\mathrm{i} \mathrm{n} \mathrm{v} \left( \mathbf{L}^{\left( a,b,c,d\right)  }\right)  =\left( \mathbf{L}^{\left( a,b,c,d\right)  }\right)^{\mathbf{H}}  =\mathbf{Q\Lambda}_{-\alpha } \mathbf{Q^{T}}_{\sigma }\mathbf{\Lambda}_{-\xi } ,
\end{equation}
where $\mathrm{i} \mathrm{n} \mathrm{v} \left( \cdot \right) $ represents the inverse matrix, and $ \left( \cdot \right)^{\mathbf{H}} $represents the conjugation transpose. An alternative to the IGLCT is to recalculate the parameters of another GLCT whose matrix is the inverse of the previous GLCT matrix:
\begin{equation}
	\mathrm{i} \mathrm{n} \mathrm{v} \left( \mathbf{L}^{\left( a,b,c,d\right)  }\right) =\mathbf{L}^{\left( d,-b,-c,a\right)  }=\mathbf{\Lambda}_{\hat{\xi } } \mathbf{Q}_{\hat{\sigma } }\mathbf{\Lambda}_{\hat{\alpha } } \mathbf{Q^{T}}.
\end{equation}

\item Reduction of the DLCT on the cycle graphs \\
It has been discussed before that when $\mathbf{A} = \mathbf{C}$, the GFT matrix and the DFT matrix are identical, so that the GLCT matrix is the same as the DLCT matrix in the cyclic graph, where the scaling parameter $\sigma=1$.

\item Convolution theorem \\
The convolution in the domain of the linear canonical transform on a graph can be given by
\begin{equation}
	\begin{aligned}\left( f\ast g\right)  \left[ n\right]  &:=\sum^{N-1}_{m=0} \hat{f} \left[ m\right]  \hat{g} \left[ m\right]  L^{\left( d,-b,-c,a\right)  }\left[ n,m\right]  \\ \text{or} \  \  \mathbf{f} \ast \mathbf{g} &:=\mathbf{L}^{\left( d,-b,-c,a\right)  } \left( \hat{\mathbf{f}} \odot  \hat{\mathbf{g}}\right)  ,\end{aligned} 
\end{equation}
where, the $\odot $ denotes the Hadamard product of the matrices. The convolution in the vertex domain is a Hadamard product in the domain of the graph linear canonical transform.
\begin{equation}
	\begin{aligned}\widehat{\mathbf{f} \ast \mathbf{g} } &=\widehat{\mathbf{L}^{\left(  d,-b,-c,a\right)  } \left(  \hat{\mathbf{f} } \odot \hat{\mathbf{g} } \right) } \\ &=\mathbf{L}^{\left( a,b,c,d\right)  } \cdot \mathbf{L}^{\left( d,-b,-c,a\right)  } \left( \hat{\mathbf{f} } \odot \hat{\mathbf{g} } \right)  \\ &=\hat{\mathbf{f} } \odot \hat{\mathbf{g} } .\end{aligned} 
\end{equation}

\item Translation property \\
First, we define a $\delta$ function:
\begin{equation}
	\delta_{i} \left[ n\right]  =\begin{cases}1,&n=i\\ 0,&\text{otherwise} \end{cases} .
\end{equation}

The translation operator $\mathcal{T}_{i}$ can be obtained by convolution with the $\delta_{i} $ function:
\begin{equation}
	\begin{aligned}\mathcal{T}_{i} \mathbf{f}:=\sqrt{N} \mathbf{f}\ast \delta_{i} &=\sqrt{N} \mathbf{L}^{\left( d,-b,-c,a\right)  }\left( \hat{\mathbf{f}} \odot \hat{\delta }_{i} \right)  \\ &=\sqrt{N} \mathbf{L}^{\left( d,-b,-c,a\right)  }\left( \hat{\mathbf{f}} \odot \mathbf{L}^{\left( a,b,c,d\right)  }_{i}\right),  \end{aligned} 
\end{equation}
where $\mathbf{L}^{\left( a,b,c,d\right)  }_{i}$ denotes the $i$th column of the matrix $\mathbf{L}^{\left( a,b,c,d\right)  }$  . Then its GLCT is calculated to be
\begin{equation}
	\begin{aligned}\mathcal{L}^{\left( a,b,c,d\right)  } \left( \mathcal{T}_{i}\mathbf{f}\right)  &=\sqrt{N} \mathcal{L}^{\left( a,b,c,d\right)  }\mathbf{L}^{\left( d,-b,-c,a\right)  } \left( \hat{\mathbf{f}} \odot \mathbf{L}^{\left( a,b,c,d\right)  }_{i} \right)  \\ &=\sqrt{N} \hat{\mathbf{f}} \odot \mathbf{L}^{\left( a,b,c,d\right)  }_{i} .\end{aligned} 
\end{equation}

\item Unitary \\
The adjacency matrix $\mathbf{A}$ of the real edge-weighted undirected graph is a real symmetric matrix, so $\mathbf{A}$ can be orthogonally diagonalized:
\begin{equation}
	\mathbf{A}=\mathbf{V\Lambda_{(A)} V}^{-1}=\mathbf{V\Lambda_{(A)} V^{T}},
\end{equation}
where $\mathbf{L}=\mathbf{V^{T}}$ is an orthogonal matrix, so it can be decomposed unitarily as:
\begin{equation}
	\mathbf{L} =\mathbf{V^{T}} =\mathbf{\Lambda_{0}Q_{1}\Lambda} \mathbf{Q}^{-1}=\mathbf{\Lambda_{0}Q_{1}\Lambda} \mathbf{Q^{H}} .
\end{equation}

Since the modulus of the eigenvalues of the orthogonal matrix is 1, the diagonal elements of the diagonal matrix $\mathbf{\Lambda}$ can be written in the form of $\exp \left( j\theta \right)  $. Thus, by the definition of
\begin{equation}
	\mathbf{L}^{\left( a,b,c,d\right)  } =\mathbf{\Lambda }_{\xi } \mathbf{Q}_{\sigma } \mathbf{\Lambda }^{\alpha } \mathbf{Q^{H}} .
\end{equation}

We consider multiplying $\mathbf{L}^{\left( a,b,c,d\right)  } $ with its conjugate transpose, 
\begin{equation}
	\begin{aligned}\mathbf{L}^{\left( a,b,c,d\right)  } \left( \mathbf{L}^{\left( a,b,c,d\right)  } \right)^{\mathbf{H} }  &=\left( \mathbf{\Lambda }_{\xi } \mathbf{Q}_{\sigma } \mathbf{\Lambda }^{\alpha } \mathbf{Q^{H}} \right)  \left( \mathbf{\Lambda }_{\xi } \mathbf{Q}_{\sigma } \mathbf{\Lambda }^{\alpha } \mathbf{Q^{H}} \right)^{\mathbf{H} }  \\ &=\mathbf{\Lambda }_{\xi } \mathbf{Q}_{\sigma } \mathbf{\Lambda }^{\alpha } \mathbf{Q^{H}} \mathbf{Q\Lambda }^{-\alpha } \mathbf{Q^{T}}_{\sigma } \mathbf{\Lambda }_{-\xi } \\ &=\mathbf{I} \  , \end{aligned} 
\end{equation}
therefore, $\mathbf{L}^{\left( a,b,c,d\right)  }$ is also a unitary matrix.
\end{enumerate}

\section{Simulation}
To verify the validity, rationality and characteristics of the GLCT, we provide two examples to illustrate the GLCT. The calculations are based on the Graph Signal Processing toolbox (GSPBox) \cite{34} in MATLAB.
\subsection{The Cycle Diagram}
Figure 4 shows examples of bipolar rectangular signals on the cycle diagram \cite{33} to illustrate the GLCT. Figure 5 shows a comparison of the imaginary and real graphs for the cycle diagram under different $\xi$. We can understand the transformation process from vertex domain to the frequency domain by observing the GLCT of different $\xi, \sigma, \alpha$.

Specifically, the first example shows a GLCT degraded to a traditional DLCT on a cycle diagram, where $\sigma \equiv 1$. When $\xi$ is a constant, with the increase in $\alpha$, the eigenvalue (frequency) of the signal increases. When $\alpha$ is constant, the real part of the signal becomes slower as $\xi$ increases, while the imaginary part becomes steeper as $\xi$ increases.

\begin{figure}[ht]
	\begin{center}
		\includegraphics[scale=0.43]{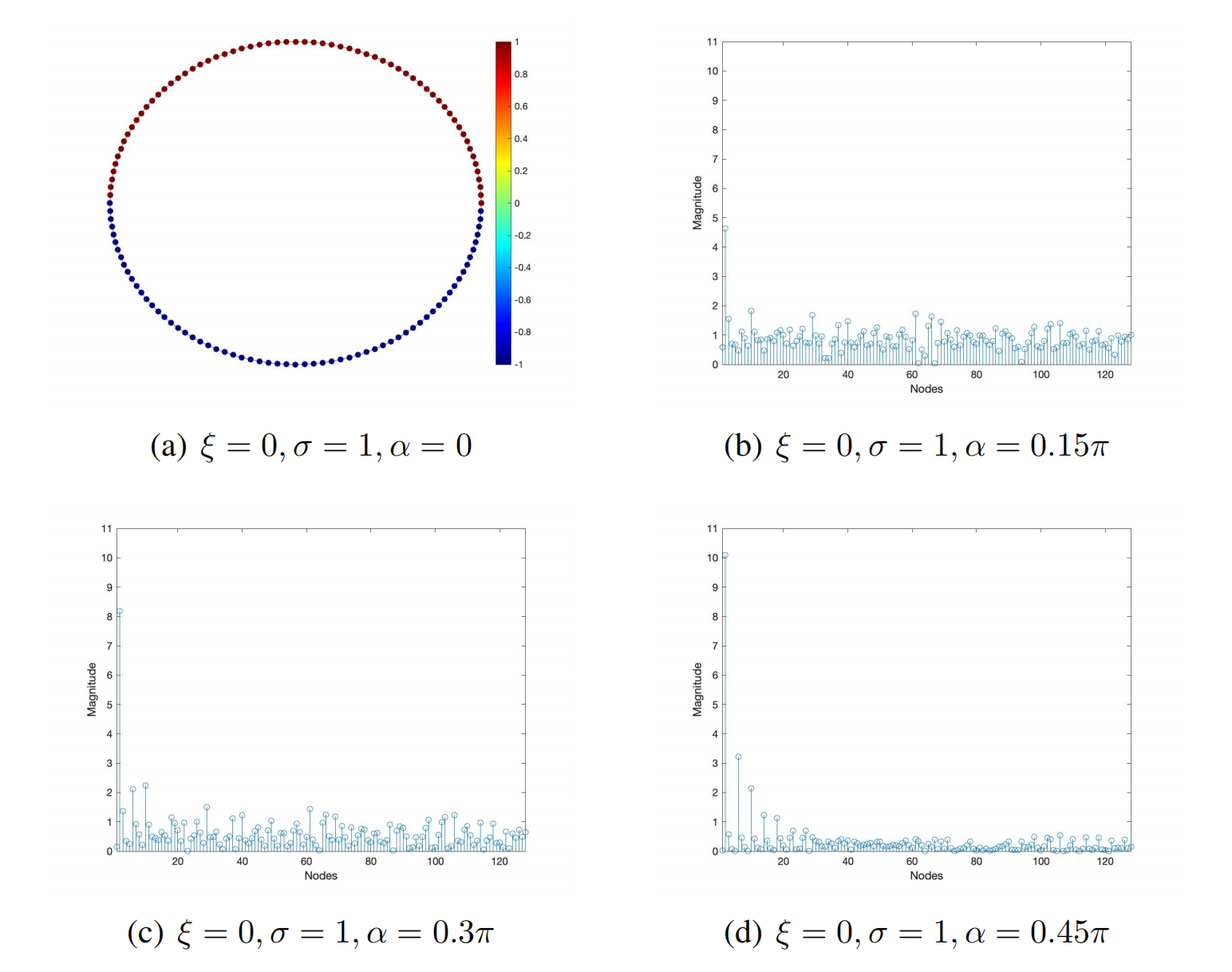}
	\end{center}
	\caption{The GLCT of the bipolar rectangular signal on the cycle graph with different $\alpha$.}
	\vspace*{-3pt}
\end{figure}

\begin{figure}[ht]
	\begin{center}
	\includegraphics[scale=0.43]{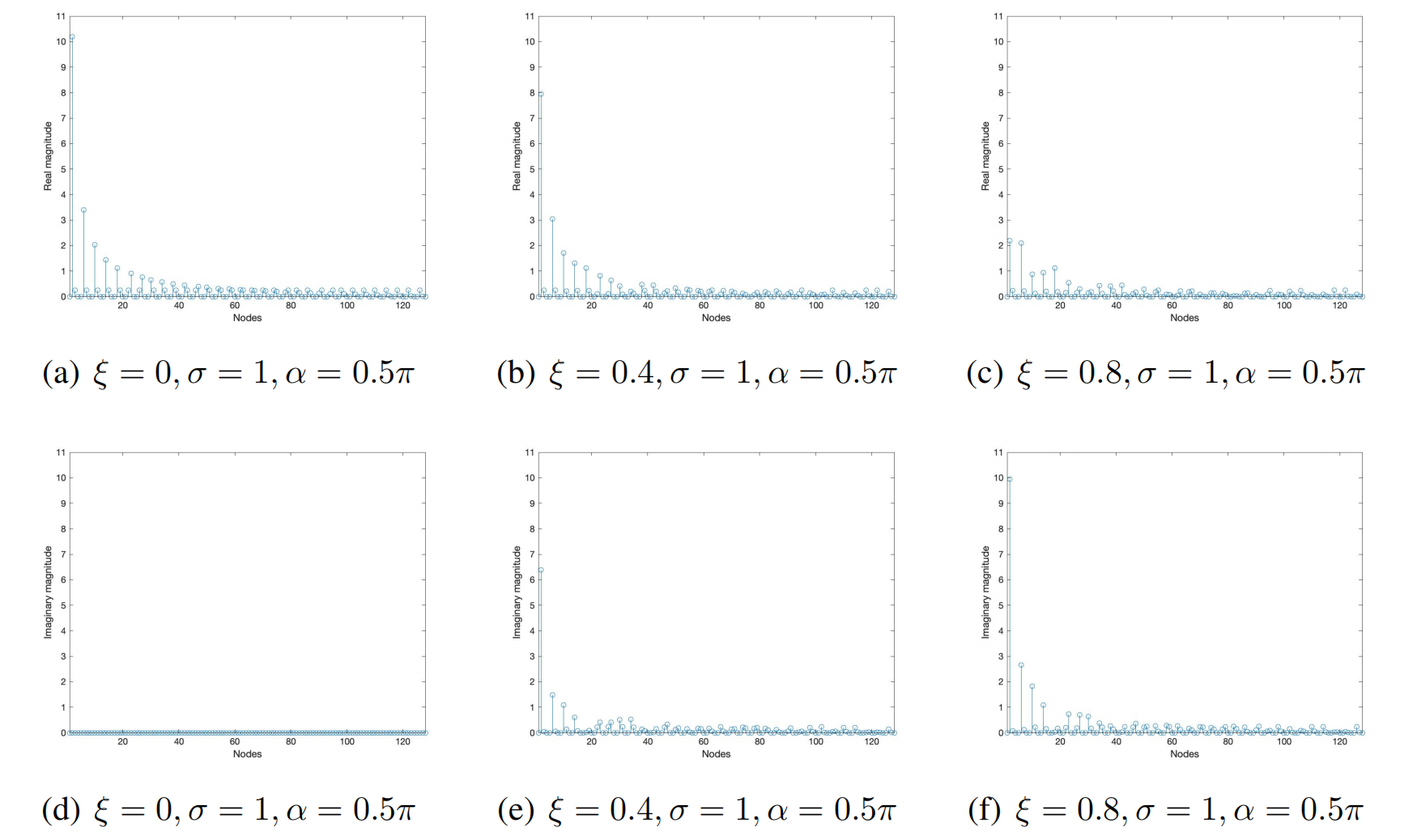}
\end{center}
\caption{Comparison of the GLCT real and imaginary values of the bipolar rectangular signals on cycle graphs with different $\xi$.}
\vspace*{-3pt}
\end{figure}

\newpage

\subsection{The Minnesota Road Diagram}
Figure 6 depicts examples of bipolar rectangular signals on the Minnesota road diagram \cite{35} to illustrate GLCT. Figure 7 shows a comparison of the imaginary and real graphs for the Minnesota road diagram under different $\xi$. We can understand the transformation process from the vertex domain to the frequency domain by observing the GLCT of different $\xi, \sigma,$and $\alpha$.

These properties can also be seen from the second example, and when $\xi,$ and $\alpha$ are constant, as $\sigma$ increases, the eigenvalues of the graph signal stabilize. 

\begin{figure}[ht]
	\begin{center}
		\includegraphics[scale=0.43]{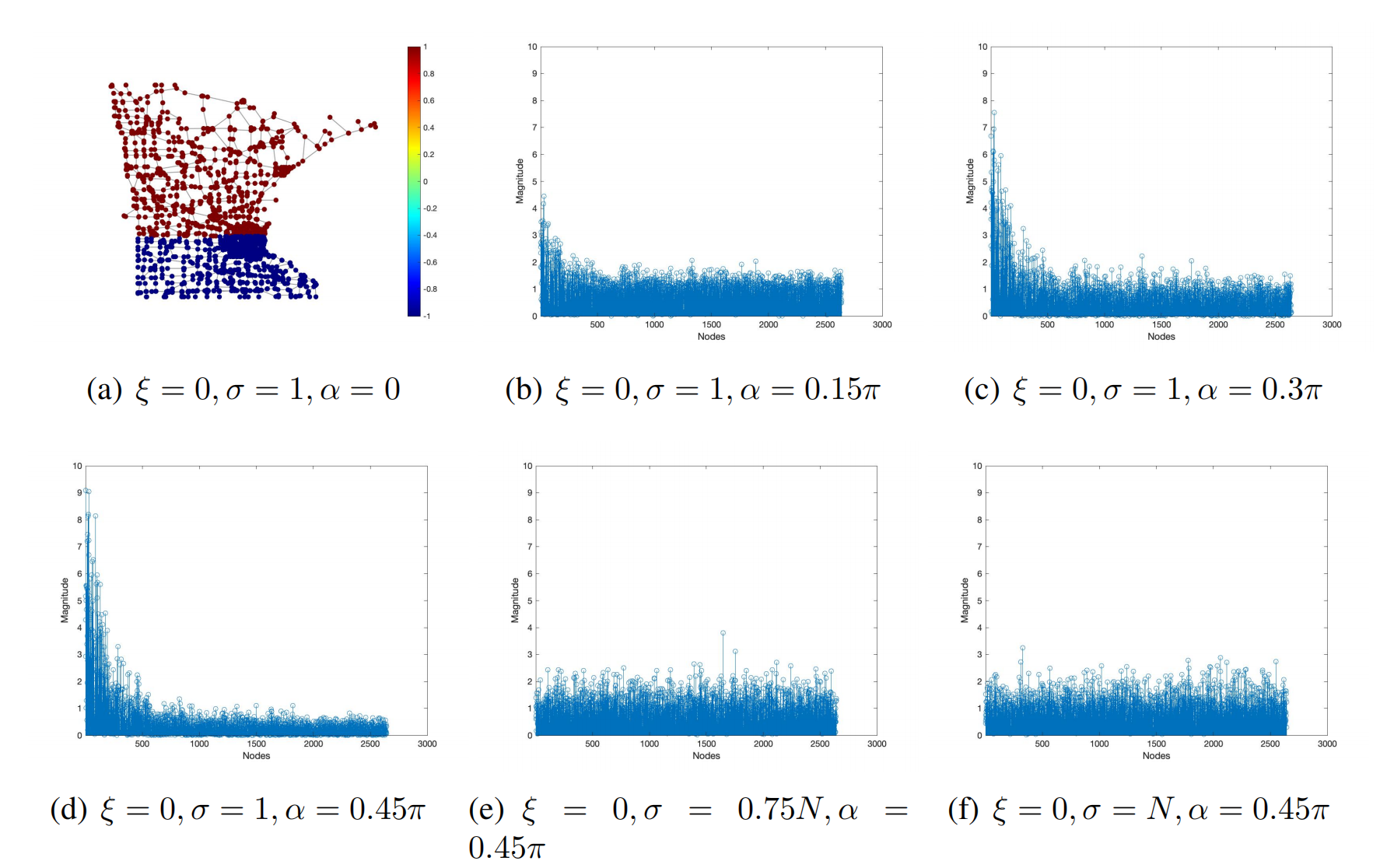}
	\end{center}
\caption{The GLCT of  the bipolar rectangular signal on the Minnesota road graph with different $\sigma, \alpha$.}
	\vspace*{-3pt}
\end{figure}

\begin{figure}[ht]
\begin{center}
	\includegraphics[scale=0.43]{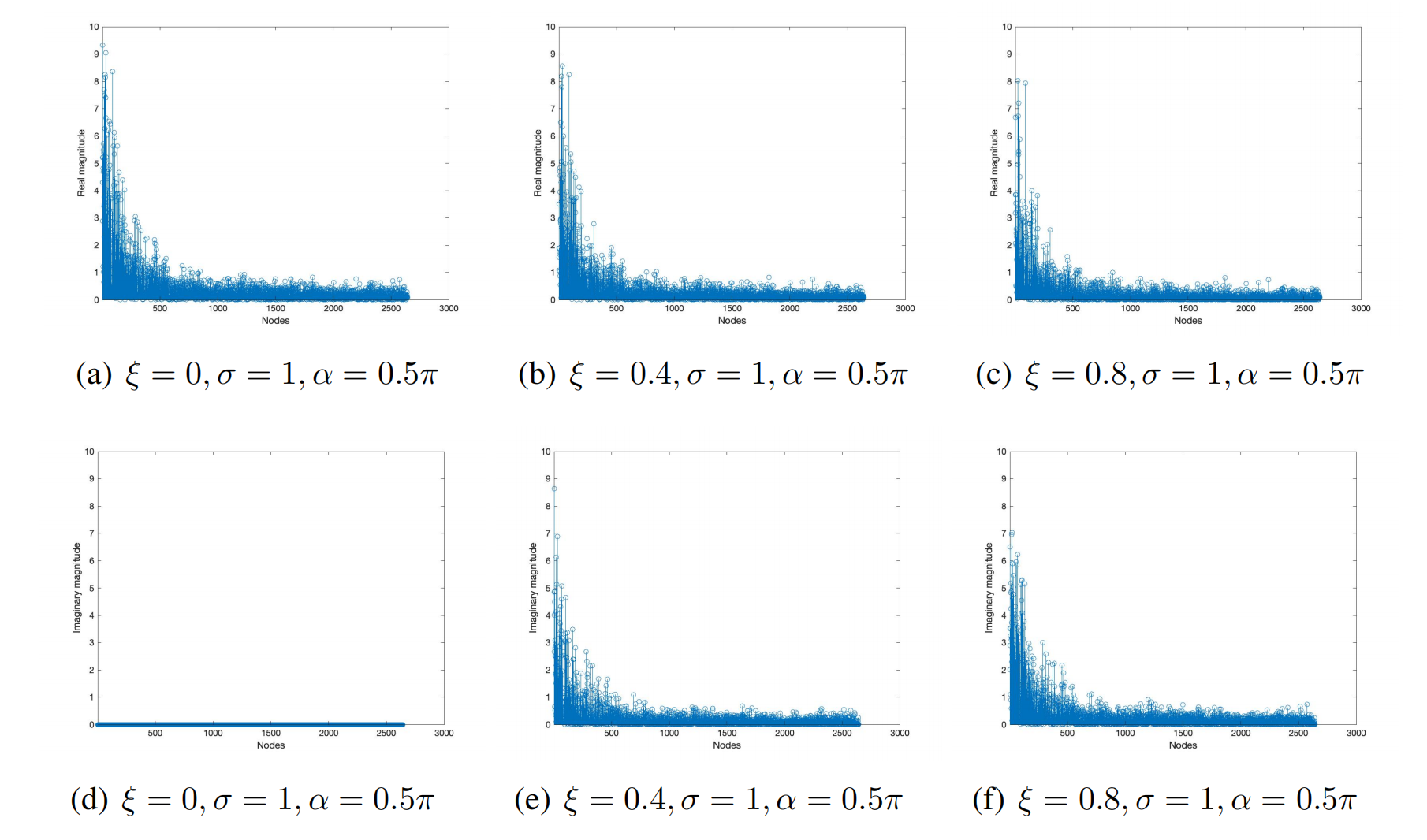}
\end{center}
\caption{Comparison of the GLCT real and imaginary values of the bipolar rectangular signals on the Minnesota road graph with different $\xi$}
	\vspace*{-3pt}
\end{figure}

\newpage

\section{ Conclusions}
To solve the problems of insufficient degrees of freedom, inadequate flexibility, and underutilized parameters that cannot be solved by the GFrFT, a generalization of the DLCT, which is called linear canonical transform on graphs, abbreviated as GLCT, is proposed in this paper. The transform depends on the parameters $\xi, \sigma$ and $\alpha$, which can solve the above problems very well. Specifically, GLCT can be reduced to GTrFT, GFT and the identity operator, and when the adjacency matrix $\mathbf{A=C}$, GLCT is reduced to the traditional DLCT. Two successive GLCTs are equivalent to another GLCT. The innovations of this paper are as follows. We provide the definition of a chirp signal on the graph, use the new eigendecomposition method (CDDHFs) to define the linear canonical transform on the graph, and carry out emulation experiments. Finally, the eigendecomposition method of the graph signal adjacency matrix used in this paper is not unique, nor is the matrix used in the chirp multiplication and scaling stages, and the current research may open up new areas in the field of graph signal processing and lead to further research. It is hoped that there will be more transformation methods and matrix forms in future research and that useful processing techniques and specific applications can be developed in the field of linear canonical transforms on graphs.

\section*{Declaration of competing interest}
The authors declare that they have no known competing financial interests or
personal relationships that could have appeared to influence the work reported
in this paper.



\end{document}